\documentclass[12pt,preprint]{aastex}
\newcommand{\mincir}{\raise
-3.truept\hbox{\rlap{\hbox{$\sim$}}\raise4.truept\hbox{$<$}\ }}
\newcommand{\magcir}{\raise
-3.truept\hbox{\rlap{\hbox{$\sim$}}\raise4.truept\hbox{$>$}\ }}
\newcommand{\minmag}{\raise
-3.truept\hbox{\rlap{\hbox{$<$}}\raise5.truept\hbox{$<$}\ }}
\newcommand{\be}{\begin{equation}}
\newcommand{\ee}{\end{equation}}
\newcommand{\ba}{\begin{eqnarray}}
\newcommand{\ea}{\end{eqnarray}}
\newcommand{\brr}{\begin{array}}
\newcommand{\err}{\end{array}}
\newcommand{\bc}{\begin{center}}
\newcommand{\ec}{\end{center}}

\shorttitle{Galaxy Bias in Quintessence Models}
\shortauthors{Basilakos \& Plionis}

\begin{document}

\title{Galaxy Bias in Quintessence Cosmological Models}

\author{S. Basilakos$^1$ \& M. Plionis$^{2,1}$}
\affil{$^1$ Institute of Astronomy \& Astrophysics, National Observatory of
Athens, I.Metaxa \& B.Pavlou, P.Penteli 152 36, Athens, Greece
}
\affil{$^2$ Instituto Nacional de Astrofisica, Optica y Electronica (INAOE)
Apartado Postal 51 y 216, 72000, Puebla, Pue., Mexico}

\section*{Abstract}
We derive the evolution of the linear bias factor, $b(z)$,
in cosmological models driven by an exotic fluid with an equation
of state: $p_{x}=w\rho_{x}$, where $-1\le w<0$ 
(quintessence). Our aim
is to put constrains on different cosmological and biasing models
by combining the recent observational clustering results of optical 
({\em 2dF})
galaxies (Hawkings et al.) with those predicted by the models.
We find that our bias model when fitted to the {\em 2dF} clustering
results predicts different bias evolution for different values of $w$.
The models that provide the weak biasing ($b_{\circ} \sim 1.1$) 
of optical galaxies found in many recent observational studies
are flat, $\Omega_{\rm m}=0.3$ with $w\le -0.9$.
These models however, predict a weak redshift evolution of $b(z)$, not
corroborated by N-body simulations.

{\bf Keywords:} cosmology: theory - large-scale structure of universe 

\newpage

\section{Introduction}
The statistical properties of the distribution of matter on 
large scales, based on different extragalactic objects, 
can provide important constrains on models of cosmic structure formation. 
However, a serious problem that hampers the straight forward use of such an
approach is our ignorance of how luminous matter trace the
underlying mass distribution. Many authors have claimed that the large scale 
clustering of different mass tracers (galaxies or clusters) 
is biased with respect to the matter distribution (cf. Kaiser 1984),
which is an essential ingredient for cold dark matter (CDM) 
models to reproduce the observed galaxy 
distribution (cf. Davis et al. 1985).
Usually, biasing is assumed to be statistical in nature by which 
galaxies and clusters are identified as
high peaks of an underlying, initially Gaussian, random density field. 

Furthermore, the bias redshift evolution
is very important in order to relate observations with models of
structure formation and it has been shown that the 
bias factor, $b(z)$, is a monotonically increasing function of redshift. 
Indeed, Mo \& White (1996) and Matarrese et al. (1997) have developed a 
model for the evolution of the correlation bias, defined as the ratio of the 
halo to the mass correlation function, the so called galaxy merging bias 
model, with $b(z) \propto (1+z)^{1.8}$. Nusser \& Davis (1994), 
Fry (1996), Tegmark \& Peebles (1998) produced a bias evolution 
model 
assuming only that the test particles fluctuation field is proportional
to that of the underlying mass and in this case, 
we have:
$b(z)=1+(b_{\circ}-1)(1+z)^{-1}$, 
where $b_{\circ}$ is the bias factor at the present time. 
Coles, Melott \& Munshi (1999) have developed a bias model within the 
hierarchical clustering paradigm which gives interesting scaling 
relations for the galaxy bias.
Recently, another approach was 
proposed by Basilakos \& Plionis (2001), in which the bias evolution was
described via the solution  of a second order differential equation, 
derived using linear perturbation theory and the basic Cosmological
equations.

From the observational point of view, recent advances in cosmology 
have strongly indicated that we are living in a 
flat, accelerating universe with low matter (baryonic and dark matter) 
density (cf. Riess et al. 1998; Perlmutter et al. 1999; de Bernadis et
al. 2000; Efstathiou et al. 2002; Persival et al. 2002, Spiergel et al
2003).
The available high quality cosmological data (Type Ia supernovae, 
CMB, etc.) are well fitted by an emerging ``standard model'', which
contains cold dark matter (CDM), to explain clustering, 
and an extra component with negative pressure, usually named 
``quintessence'',  which is in agreement with 
the inflationary flatness prediction ($\Omega_{\rm tot}=1$) as well.

In this paper we derive the redshift evolution of the linear 
scale-independent bias parameter for the ``quintessence'' models, 
and additionally we attempt to put constraints on 
the different cosmological models using the recent 
clustering results (Hawkins et al. 2002) of optical 
galaxies from the {\em 2dF} survey. We compare the observed clustering
of {\em 2dF} galaxies with that expected in 10 spatially flat 
cosmological models with dark energy and $\Omega_{\rm m}=0.3$. 
It is worth noting that the galaxy bias has been found to be 
sensitive to the equation of state (Munshi, Porciani \& Wang 2003).


\section{Linear Bias Evolution Model}
In this section we briefly describe our linear bias evolution model 
(for more details see Basilakos \& Plionis 2001) which is based
on linear perturbation theory in the matter dominated epoch (cf. Peebles 1993).
We assume that the mass tracer population is conserved
in time, ie., that the effects of non-linear gravity and 
hydrodynamics (merging, feedback mechanisms etc) can be ignored 
(cf. Fry 1996; Tegmark \& Peebles 1998; 
Catelan et al. 1998), and using the linear perturbation theory 
we can obtain a second order differential equation, which describes the 
evolution of the linear bias factor, $b$, between the background
matter and a mass-tracer fluctuation field: 
\be\label{eq:11}
\ddot{b}D(t)+2 \left[\dot{D}(t)+H(t)D(t) \right]\dot{b}+4\pi G \rho_{\rm m}
D(t) b = 4\pi G \rho_{\rm m} D(t) \;\;\;,
\ee
where $D(t)$ is the linear growing mode, useful expressions 
of which can be found for the $\Lambda$CDM model in
Peebles et al. (1993) and for the QCDM models in Wang \& Steinhardt 
(1998).
The solutions of this second
order differential equation provides our bias evolution model. The
solution for the different cosmological model enter through the
different behaviour of $D(t)$ and $H(t)$.
To see the relevance and difference of our bias evolution model to
the {\em Galaxy Conserving} one, which is based on similar
assumptions, we will derive somewhat
differently eq.(\ref{eq:11}), starting from the continuity
equation, which is the starting point of the {\em Galaxy Conserving} model. 
If the galaxies and the underlying matter share the same velocity field then:
$\dot{\delta} + \nabla u \simeq 0$ and $\dot{\delta}_g +
\nabla u \simeq 0$, 
from which we get $ \dot{\delta} - \dot{\delta}_g=0$. Now since we assume 
linear biasing, we have $\delta_g=b \delta$ and using $y=b-1$, we get
that $d(y \delta)/dt=0$. Differentiating twice we then have:  
$\ddot{y} \delta + 2 \dot{y} \dot{\delta} + y \ddot{\delta} =0$.
Solving for $y \ddot{\delta}$, using the fact that $y
\dot{\delta} = -\dot{y}\delta$ and utilizing the differential
time-evolution equation of $\delta$ (cf. Peebles 1993) we finally
obtain:
\be
\ddot{y}\delta + 2(\dot{\delta} + H \delta) \dot{y} + 4 \pi G
\rho \delta y =0
\ee
which is the corresponding equation \ref{eq:11}.

Here we extend our previous bias evolution
solutions of Basilakos \& Plionis (2001)
to flat Friedmann-Robertson Walker Cold Dark Matter
(CDM) type models driven by non-relativistic
matter and having an exotic fluid (quintessence) with equation of state:
\be\label{eq:eos}
p_{x}=w \rho_{x} \;\;\; (-1\leq w <0) \;\;\;.
\ee

Following the notation of Basilakos \& Plionis (2001), 
we present the basic steps of our procedure. 
In order to transform (\ref{eq:11}) 
from time to redshift we utilize the following expressions:
\be\label{eq:5}
\frac{dt}{dz}=-\frac{1}{H_{\circ}E(z)(1+z)} \;\;\;,\;\;\;
E(z)=\left[ \Omega_{\rm m}(1+z)^{3}+\Omega_{x_o}(1+z)^{\beta}\right]^{1/2}
\ee
while $\Omega_{\rm m}= 8\pi G \rho_{o}/3H_{o}^{2}$ 
(density parameter), $\Omega_{x_o}= 8\pi G \rho_{x_o}/3H_{o}^{2}$ 
(dark energy parameter) at the present time, which satisfy
$\Omega_{\rm m}+\Omega_{x_o}=1$, and $\beta=3(1+w)$ with $0\le \beta <3$. 
Taking into account the latter transformations, the 
basic differential equation for the evolution of the linear
bias parameter takes the following form:
\be\label{eq:gen2}
\frac{{\rm d}^{2} b}{{\rm d} z^{2}}-P(z)\frac{{\rm d} b}{{\rm d} z}+
Q(z)b=Q(z) \;\; 
\ee
with relevant factors, 
\be\label{eq:ff1}
P(z)=\frac{1}{1+z} - \frac{1}{E(z)}\frac{{\rm d}E(z)}{{\rm d}z}
-\frac{2}{D(z)}\frac{{\rm d}D(z)}{{\rm d}z}
\ee
and
\be\label{eq:ff2}
Q(z)=\frac{3\Omega_{\rm m} (1+z)}{2E^{2}(z)} \;\; .
\ee

Therefore, the general bias solution 
for all of the flat cosmological models is:    
\be
b(z)= {\cal A} E(z)+{\cal C} E(z)I(z)+1 \;\;\;, 
\ee
where
\be\label{eq:88} 
I(z)=\int \frac{(1+z)^{3}}{E(z)^{3}}  {\rm d}z\;\;\;.
\ee
The integral of equation (\ref{eq:88}) is elliptic and therefore
its solution, in the redshift range $[z,+\infty)$, can be expressed 
as a hyper-geometric function. Our final general solution is:
\be\label{eq:89}  
b(z)-1={\cal A} E(z)+\frac{2{\cal C}}{\Omega_{\rm m}^{3/2}} E(z) 
(1+z)^{\frac{\beta-3}{6-2\beta}} 
F\left[\frac{1}{6-2\beta},\frac{3}{2},\frac{7-2\beta}{6-2\beta},-
\frac{\Omega_{x}}
{\Omega_{\rm m} (1+z)^{3-\beta}} \right] \;\; . 
\ee
Note that the first term, which is the dominant one, has
an approximate 
redshift dependence $\sim (1+z)^{3/2}$ while the second has $\sim (1+z)$.
Also note that for $w \longrightarrow -1$  
the above general bias solution tends to the $\Lambda$CDM case, as it should 
(eq.39 of Basilakos \& Plionis 2001).
It is quite instructive to see the solution for the case of $w=-1/3$,
in which case  
the functional form of the bias evolution is independent of the 
hyper-geometric function\footnote{Where we have 
used $\Omega_{x}=1-\Omega_{\rm m}$ and 
$F[\alpha,b,b,t]=(1-t)^{-\alpha}$.}. We have that:
\be
b(z)={\cal A} (1+z)(1+\Omega_{\rm m}z)^{1/2}
+\frac{2{\cal C}}{\Omega_{\rm m}}(1+z)+1 \;\;\;.
\ee
We can easily verify that for $\Omega_{\rm m}=1$ this reduces to the
Einstein-de Sitter case (eq. 29 of Basilakos \& Plionis 2001), as it should.
 
\section{Estimating the Bias from Galaxy and Mass correlations}
Since our approach gives a family of bias
curves, due to the fact that it has two unknown parameters, 
(the integration constants ${\cal A},{\cal C}$) and
in order to obtain partial solutions for $b(z)$ we need to estimate the 
values of these constants. In Basilakos \& Plionis (2001) we have
compared our bias evolution model with the halo merging model (cf. Mo
\& White 1996; Mataresse et al. 1997) as well as with 
different N-body results and
found a very good consistency, once we fitted the integration 
constants ${\cal A},{\cal C}$ by evaluating our model to two
different epochs. We further compare in figure 1 
our solution for the $\Lambda$CDM case (see its parameters further below), 
evaluated at $z=0$ and $z=3$ using the HDF results (Arnouts et al. 2002; Malioccietti
1999, with the Mataresse et al (1997) model. It is quite evident that our
model fits better the $z$-dependance of the observational HDF galaxy bias.

Although, this comparison gives consistent results of
the functional form of our solution with available data and 
theoretical models (see Basilakos \& Plionis 2001), it does not
test directly whether our model, once calibrated observationally, provides
a consistent model at high $z$'s. Therefore, in order to test it, we evaluate 
the constants ${\cal A},{\cal C}$ by fitting our bias model to the the recent 
{\em 2dF} galaxy clustering results of Hawkins et al. (2002) who used  
$\sim$200000 galaxies observed in the {\em 2dF} survey to derive
their spatial correlation function. 

We use the standard theoretical approach  
to estimate the two point spatial correlation 
function, using our model for the bias evolution in the
different spatially flat cosmological models.
We quantify the evolution of clustering with epoch, 
writing the spatial galaxy correlation function 
as $\xi_{\rm model}(r,z)=\xi_{\rm mass}(r)R(z)$,
with $R(z)=D^{2}(z)b^{2}(z)$, where
the function $R(z)$ characterizes the clustering evolution
with epoch (see Basilakos 2001 and references therein). 
While the $\xi_{\rm mass}(r)$ is the Fourier
transform of the spatial power spectrum $P(k)$:
\be
\xi_{\rm mass}(r)=\frac{1}{2\pi^{2}}\int_{0}^{\infty} k^{2}P(k) 
\frac{{\rm sin}(kr)}{kr}{\rm d}k \;\;,
\ee
where $k$ is the comoving wavenumber. Note that we also use the
non-linear corrections introduced by Peacock \& Dodds (1994).
As for the power spectrum of our CDM models, we take 
$P(k) \approx k^{n}T^{2}(k)$ with scale-invariant ($n=1$) primeval 
inflationary fluctuations and $T(k)$ the CDM transfer function.
In particular, we use the transfer function parameterization as in
Bardeen et al. (1986), with the corrections given approximately
by Sugiyama's formula (Sugiyama 1995). 

In the present analysis we consider flat models with cosmological 
parameters that fit the majority of observations, ie.,
$\Omega_{\rm m}+\Omega_{x_o}=1$, $\Omega_{\rm m}=0.3$, 
$H_{\circ}=100h $km s$^{-1}$ Mpc$^{-1}$
with $h\simeq 0.6 - 0.7$ (Freedman et al. 2001; 
Peebles and Ratra 2002 and references therein) and baryonic density 
parameter $\Omega_{\rm b} h^2 \simeq 0.02$ (cf. 
Olive, Steigman \& Walker 2000; Kirkman et al 2003).
We investigate such spatially flat cosmological models 
with negative pressure for different values of $w$ (see eq. \ref{eq:eos}).
To this end, all the cosmological models 
are normalized to have fluctuation amplitude in 8 $h^{-1}$Mpc scale of
$\sigma_{8}=0.50 (\pm0.1) \Omega_{\rm m}^{-\gamma}$ 
(Wang \& Steinhardt 1998) with 
$\gamma=0.21-0.22w+0.33\Omega_{\rm m}$. 

In order to quantify the integration constants $(\cal{A},\cal{C})$
we perform a standard $\chi^{2}$ 
minimization procedure between the measured 
correlation function for the {\em 2dF} galaxies (Hawkins et al. 2002) 
with those expected in our spatially flat cosmological models,
\be
\chi^{2}=\sum_{i=1}^{n} \left[ \frac{\xi_{2df}^{i}(r)-\xi_{\rm model}^{i}(r)}
{\sigma^{i}}\right]^{2} \;\;.
\ee 
where $\sigma$ is the observed correlation function uncertainty.

In figure 2 we present 
the fit to the data of two models with values of $w$ at the opposite
end of the range explored. Hawkins et al (2002) found  that the 
{\em 2dF} clustering behaviour on small scales 
is represented by a power law, with correlation length
$r_{\circ}\simeq 5.05h^{-1}$Mpc and slope $\gamma \simeq 1.67$.
Due to the interplay between the three unknown (${\cal A},
{\cal C}, b_\circ$) all of the models (for different values of the
constants) fit at a high significance level 
($P_{\chi^{2}} \approx 0.7$) the {\em 2dF} correlation function.
 In Table 1 we list the results of the fits for all models, ie., the
integration constants and the value of the optical bias, $b_{\circ}$, at the 
present time, derived from our model (eq. \ref{eq:89}).
It is interesting that for the {\em standard} $\Lambda$CDM model 
($w=-1$, $\Omega_m=0.3$ and $h=0.7$) the $\sim
{\cal C} (1+z)$ factor of the solution is $\simeq 0$, which means that
the bias evolution is a very weak function of $z$, approximated by: 
$b(z)-1 \approx \frac{1}{33} (1+z)^{3/2}$.
  
It is also very interesting that for the $\Lambda$CDM and the
$w=-0.9$ QCDM models (with $h=0.7$), the corresponding 
bias at the present time is $b_{\circ}\simeq 1.05$ and  
$b_{\circ}\simeq 1.11$, respectively, in very good agreement
with the values $b_{\circ}\simeq 1.04$ and 1.11
derived by Verde et al. (2002) and Lahav et al. (2002), respectively.
When we use $h=0.6$,
only the $\Lambda$CDM model provides a value for the present bias which is 
only roughly consistent with other observational results
($b_{\circ}\simeq 1.17$). 
The present time bias, $b_\circ$,
as a function of $w$, is well fitted by a power law having the form:
$b_{\circ}\simeq [1.93-1.20h](\pm 0.019)\times (-w)^{-0.411(\pm 0.018)}$.

Having derived the values of the constants ${\cal A}, {\cal C}$ and
the zero-point bias, $b_\circ$, from the fit to the {\em 2dF} galaxy
clustering pattern, we can now investigate the bias evolution in the
different cosmological models. In figure 3 we present the function
$b(z)$ for different values of $w$. 
The biasing is a monotonically increasing function of redshift 
with its evolution being significantly stronger for lower values of
$w$. Galaxy clustering in quintessence models evolves
more rapidly than in the $\Lambda$CDM model.
It is also clear that for the preferred cosmological models 
(with $w\le -0.9$ and $h=0.7$)
the optical galaxies are only weakly biased, even at high redshifts.

In figure 3 we also 
compare our evolution model with the {\em galaxy conserving}
one (Fry 1996; Tegmark \& Peebles 1998) 
by normalizing the latter to the value of $b_\circ$
provided by our fit. It is interesting that the two models, despite their
different form, are consistent with each other. However, we note that
this is true for the $h=0.7$ case. Had we presented the $h=0.6$ case it
would have be evident that the consistency is significantly worse.
In the same figure we compare our model with the results of a $\Lambda$CDM
N-body simulation of Somerville et al (2001) (thick blue line). The
simulation galaxy bias evolves significantly more than what our model 
predicts, which
should be attributed to our assumption that the galaxy number density
is conserved in time. It is evident that merging processes,
not taken into account by our model, 
are very important in the evolution of clustering.

\section{Conclusions}
In this paper we derived analytically the evolution of linear bias $b(z)$
for cosmological models driven by an exotic fluid with an equation
of state: $p_{x}=w\rho_{x}$, where $-1\le w<0$. 
Comparing the optical {\em 2dF} galaxy correlation function (Hawkings 
et al. 2002) with the predictions of various (quintessence) models,  
we find that the flat cosmological models with $w\le -0.9$  
provide present bias values
in the range $1.05 \le b_{\circ}\le 1.11$, which is consistent with  
observational correlation results. 
However,  we find that for the above cosmological models our 
bias evolution is too weak to fit N-body results of galaxy clustering,
a fact which should be attributed to merging processes, not taken into
account in our model.

\begin{table*}
\caption[]{Our bias model normalizations and present linear bias
($b_{\circ}$) between optical galaxies and underlying matter
distribution.}
\vspace{0.5cm}

\tabcolsep 18pt
\begin{tabular}{cccccccc}
& \multicolumn{3}{c}{$H_{\circ}=70$ km s$^{-1}$Mpc$^{-1}$} & &
\multicolumn{3}{c}{$H_{\circ}=60$ km s$^{-1}$Mpc$^{-1}$} \\ \hline \hline
$w$ & ${\cal A}$ & ${\cal C}$ & $b_{\circ}$ &  & ${\cal A}$ & ${\cal C}$& 
$b_{\circ}$ \\ \hline \hline
-1.0 & 0.016& 0.004&1.05 & & 0.004& 0.020&1.17\\
-0.9 & 0.028& 0.010&1.11 & & 0.022& 0.028&1.23\\
-0.8 & 0.038& 0.018&1.18 & & 0.014& 0.038&1.31\\
-0.7 & 0.024& 0.032&1.27 & & 0.032& 0.050&1.41\\
-0.6 & 0.084& 0.040&1.37 & & 0.054& 0.064&1.53\\
-0.5 & 0.074& 0.064&1.50 & & 0.058& 0.092&1.67\\
-0.4 & 0.190& 0.078&1.67 & & 0.070& 0.130&1.85\\
-0.3 & 0.190& 0.130&1.86 & & 0.090& 0.190&2.07\\
-0.2 & 0.490& 0.150&2.10 & & 0.320& 0.250&2.34\\
-0.1 & 0.480& 0.320&2.39 & & 0.410& 0.440&2.67\\ \hline \hline
\end{tabular}
\end{table*}

\begin{figure}
\plotone{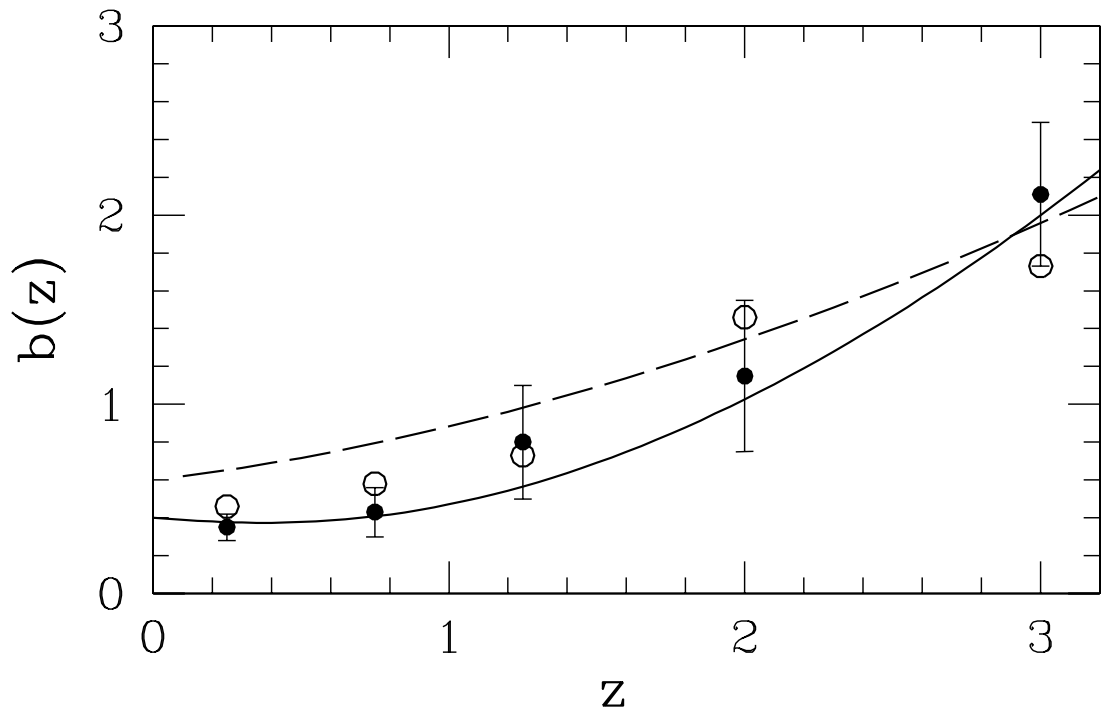}
\caption{Comparison of our bias model for the $\Lambda$CDM case (continuous line) 
with the corresponding Mataresse et al (1997) halo merging model for galaxies 
with $\sim 10^{11} \; M_{\odot}$ (dashed line). The points represent
  the HDF galaxy bias data from the HDF-north (open symboles) and
  HDF-south (filled symboles). Note that our model has been normalized
  to the data at $z=0$ and $z=3$.}
\end{figure}

\begin{figure}
\plotone{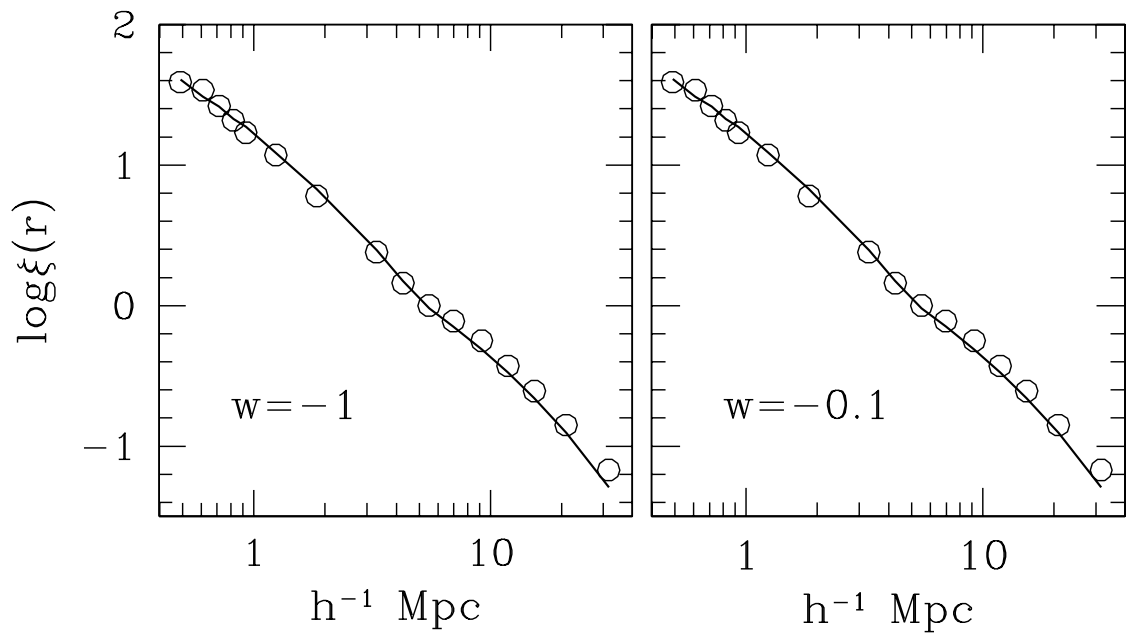}
\caption{Comparison of the {\em 2dF} correlation function
(Hawkins et al. 2002) with that predicted for the $\Lambda$CDM
($w=-1$) and an extreme quintessence model ($w=-0.1$),
using $h=0.7$. Each fit corresponds to different values of the integration 
constant ${\cal A}, {\cal C}$ and of the zero-point bias, $b_\circ$ 
(see table 1).}
\end{figure}


\begin{figure}
\plotone{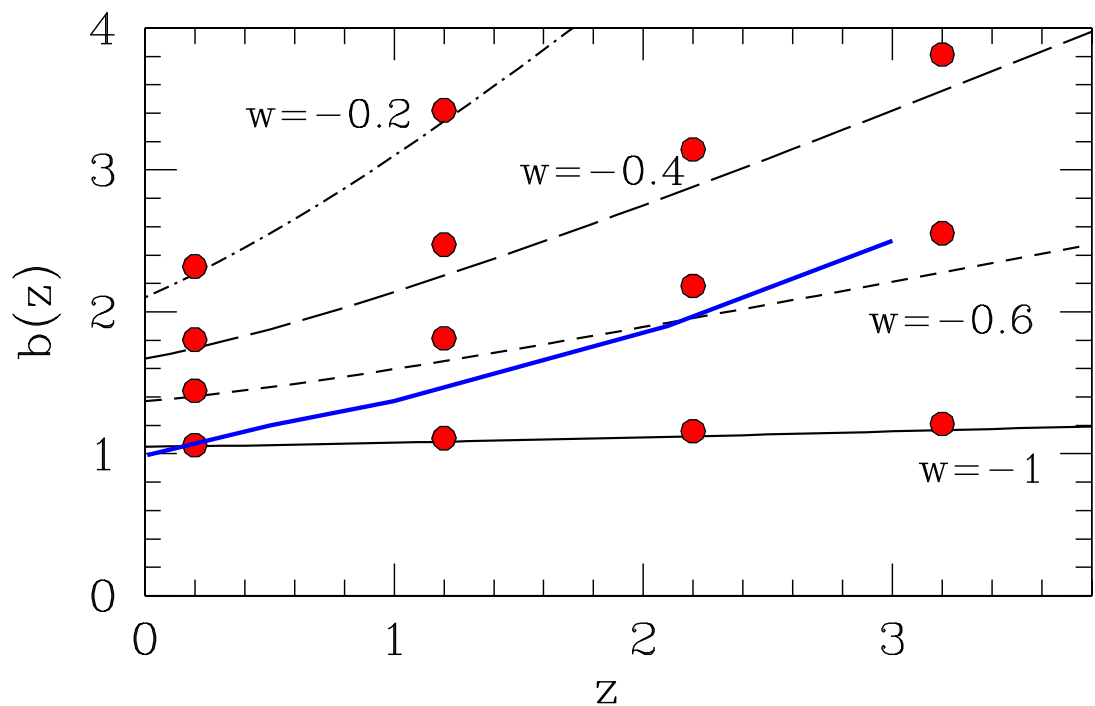}
\caption{Evolution of the linear bias factor for our model with $h=0.7$.
Different line types represent different values of
$w$ in our model, while the filled points is the Fry (1996)  
evolution model normalized to $b(0)$. The thick continuous line (blue)
represent the results of galaxies with $M_{B}-5 \log h \le -19.5$ in
the Somerville et al (2001) $\Lambda$CDM simulation while the star
like symboles represent the HDF results for the $\Lambda$CDM model of
Maliochetti \& Lahav (1998).} 
\end{figure}

\end{document}